\documentclass[11pt,a4paper]{article}

\usepackage[utf8]{inputenc}
\usepackage[T1]{fontenc}
\usepackage[english]{babel}
\usepackage{lmodern}
\usepackage[a4paper,margin=2.5cm]{geometry}
\usepackage{parskip}
\usepackage{microtype}
\usepackage{enumitem}
\usepackage[colorlinks=true,allcolors=blue,pdfusetitle]{hyperref}
\usepackage{url}
\usepackage{authblk}
\usepackage{abstract}
\usepackage{titlesec}

\titlespacing*{\section}{0pt}{1.6ex}{0.8ex}
\titlespacing*{\subsection}{0pt}{1.2ex}{0.4ex}

\title{The AI Legal Specialist: A Juridically Autonomous Professional Profile for AI Governance}

\author[1]{Nicola Fabiano\thanks{Studio Legale Fabiano, Italy. Independent Researcher on Artificial Intelligence, Data Protection, and Privacy. Expert in the EDPB's Support Pool of Experts --- Field B: Legal Expertise in New Technologies (project-based). Member, IEEE SA P7007 Working Group on Ontological Standards for Ethically Driven Robotics (Chair, subgroup ``Data Privacy and Protection''). Member, Editorial Advisory Board, Journal of Systemics, Cybernetics and Informatics (JSCI). Member, International Institute of Informatics and Systemics (IIIS). Member, International Neural Network Society (INNS). Member, United Nations University AI Network (UNU AI Network). Email: \href{mailto:info@fabiano.law}{\texttt{info@fabiano.law}}. ORCID: \href{https://orcid.org/0000-0002-8188-7656}{0000-0002-8188-7656}.}}
\affil[1]{Studio Legale Fabiano, Italy}

\date{}

\begin{document}
\maketitle

\begin{abstract}
\noindent The rapid global expansion of artificial intelligence regulation has generated, across multiple jurisdictions, a demand for legal expertise dedicated to AI that the market has addressed in a fragmented manner. Data protection officers extend their remit beyond data protection law; privacy lawyers reposition themselves toward AI; compliance officers add AI chapters to their existing manuals. This paper argues that none of these adaptive responses adequately covers the professional space opened by the emerging global AI regulatory landscape, of which the EU Artificial Intelligence Act (Regulation (EU) 2024/1689) is the most comprehensive instance, alongside the Council of Europe Framework Convention on AI, the United States executive and sectoral framework, and analogous initiatives in the United Kingdom, Canada, Brazil, China, Japan, Singapore, and beyond. A distinct professional profile is required: the AI Legal Specialist, conceived as a jurist --- understood broadly to encompass any professional with advanced legal training --- operating at the intersection of legal interpretation and AI governance. The profile is juridically autonomous: it derives its existence from the structure of regulatory obligations generated wherever AI is subject to substantive regulation, rather than from any technical standard or the extension of adjacent roles. The paper provides a juridically grounded definition of the profile, argues for its autonomy from adjacent figures and international standards, proposes a reference competence architecture aligned with the European e-Competence Framework (e-CF, EN 16234-1) as a methodological choice, and articulates the conditions for its operational measurement through key performance indicators. The contribution is intended as a foundation for international standardization of the profile and as a reference for practice, curricula, and adoption across jurisdictions.
\end{abstract}

\vspace{0.5em}

\noindent\textbf{Keywords:} AI governance; professional profile; juridical autonomy; AI Act; Council of Europe AI Convention; comparative AI law; e-Competence Framework; EN 16234-1; legal profession; AI compliance.

\vspace{0.5em}

\noindent\textit{Author's note. The structured formalization of the AI Legal Specialist profile referenced in this paper was deposited by the author with a certified timestamp (eIDAS, Regulation (EU) No 910/2014) on 11 May 2025 at the Patamu Registry (deposit no. 253295). The deposited document remains reserved at this stage; this paper presents and discusses the underlying methodology, the conceptual architecture of the profile, and its positioning within the existing standardization and professional landscape, without reproducing the specific articulation of competences and levels set out in the reference document.}

\section{Introduction}

The regulation of artificial intelligence has entered, in the period 2023--2026, a phase of structural acceleration at the global level. The European Union, with the Artificial Intelligence Act \cite{aiact2024}, has adopted the most comprehensive horizontal regulatory framework on AI to date, complemented by an ecosystem of European digital regulations comprising the GDPR \cite{gdpr2016}, the NIS2 Directive \cite{nis22022}, the Data Act \cite{dataact2023}, the Digital Services Act \cite{dsa2022}, the Digital Markets Act \cite{dma2022}, the AI Liability Directive proposal, and the Cyber Resilience Act. The Council of Europe, with the Framework Convention on Artificial Intelligence and Human Rights, Democracy and the Rule of Law \cite{coeaiconv2024}, has adopted the first binding international instrument on AI, opened for signature in September 2024 by Member States and by non-member jurisdictions including the United States, the United Kingdom, Canada, and Israel. Beyond Europe, regulatory initiatives are progressing rapidly: the United States have developed an executive and sectoral framework, the United Kingdom has articulated a pro-innovation approach distributed across existing regulators, Canada is advancing the Artificial Intelligence and Data Act proposal, Brazil is finalizing the AI Bill, China has issued comprehensive provisions on generative AI, and analogous initiatives are proceeding in Japan, Singapore, the Republic of Korea, India, and other jurisdictions. The convergence is striking: across jurisdictions and legal traditions, AI is no longer treated as a technological domain free from substantive legal obligations, but as an object of regulation whose interpretation requires sustained juridical engagement.

This convergence has generated, across multiple legal systems, a demand for legal expertise dedicated to AI. The European professional market, in the first eighteen months of the AI Act's applicability, has reacted with a pattern of informal adaptation: existing legal roles have absorbed, sometimes uncritically, the tasks associated with AI governance. Analogous adaptive dynamics can be observed in other jurisdictions where AI regulation is taking hold. This paper takes the position that, while pragmatic, such adaptation is structurally insufficient. The density of obligations introduced by emerging AI regulation, the interdisciplinary character of the subject-matter, and the pace of technological evolution \cite{fabiano2024aillm} all point to the emergence of a distinct professional figure that existing categories do not capture, regardless of the specific jurisdiction in which the obligations arise.

The profile proposed here is \textit{juridically autonomous}: it derives its existence from the structure of regulatory obligations generated wherever AI is subject to substantive regulation, rather than from the architecture of any technical standard or the extension of adjacent professional roles. This autonomy is a substantive claim, not a formal one, and it has consequences for how the profile must be articulated, trained, and recognized. The claim bears on the relation of the profile to its constitutive preconditions: without artificial intelligence as an object of regulation, there is no subject-matter for the figure; without a substantive regulatory framework on AI, there is no binding obligation from which the figure arises; without a sufficiently developed regulatory ecosystem, there is no systemic breadth that justifies a dedicated professional dimension. These three conditions, taken jointly, constitute the precondition of the existence of the profile. The profile's universality is thus conditional rather than absolute: it emerges, with the same conceptual architecture, in every jurisdiction where the three conditions are satisfied; the specific regulatory content that populates the profile's operational domain is, by contrast, jurisdiction-specific.

The paper makes three contributions. First, it offers a juridically grounded definition of the AI Legal Specialist and argues its autonomy from adjacent figures and from international standardization instruments (Section~\ref{sec:autonomy}). Second, it proposes a reference competence architecture aligned with the European e-Competence Framework, with a methodological rationale for the approach and for the reserved articulation of its specific elements (Section~\ref{sec:ecf}). Third, it articulates the conditions under which the profile can be operationalized through measurable key performance indicators (Section~\ref{sec:kpi}). Throughout, the EU AI Act is treated as the most mature instance of comprehensive AI regulation, but the analytical framework is constructed to remain valid across the multiple regulatory environments in which AI governance is being articulated.

The paper is deliberately scoped as a foundational contribution. It does not claim to exhaust the taxonomy of professional figures emerging around AI governance, nor to compete with certification schemes that address overlapping but distinct populations. It aims rather to fix, in scientific form, the conceptual architecture of a profile whose relevance will increase as AI regulation matures across jurisdictions and as international coordination instruments enter into force.

\section{Problem Setting and Methodology}

\subsection{The insufficiency of the adaptive response}

The figures most frequently invoked, across jurisdictions, to cover the new demand for AI-related legal expertise are the Data Protection Officer (DPO) or its functional equivalents (Privacy Officer, Chief Privacy Officer), the privacy lawyer, the compliance officer, and, more recently, hybrid figures circulating under labels such as ``AI Compliance Officer'' or ``AI Governance Professional''. Each of these responses, examined carefully, reveals structural limitations relative to the space that emerging AI regulation has opened.

The DPO, defined in the European context by Articles 37--39 of the GDPR and by analogous provisions in jurisdictions whose data protection law derives from or converges with the GDPR, has a mandate limited to the protection of personal data. Its extension to AI systems is legitimate only to the extent that the AI systems in question involve personal data processing; it does not cover AI systems that raise obligations under AI-specific regulation independently of any personal data dimension. The privacy lawyer, which in most jurisdictions is not a regulated profession but a market specialization, performs advisory, contractual, and litigation functions centered on data protection; AI regulation is, for this figure, an extension of the domain rather than a reconstitution of it.

The compliance officer works on organizational conformity processes; the juridical interpretation of norms is, in this role, ancillary to the operational implementation of controls. By contrast, the figure required by AI regulation has legal interpretation as its primary function, with organizational implementation as consequence. Emerging certification schemes on AI compliance and AI governance, which have proliferated in the European, North American, and international professional market during 2024--2026, typically address professionals whose background is not necessarily legal: they are accessible to risk managers, privacy professionals, and IT specialists, and the ``legal'' component of the profile is, in these schemes, one competence among others rather than the defining dimension. The absence of a juridically grounded professional profile in the international standardization landscape --- a gap also observable in the work of ISO/IEC JTC 1/SC 42 on artificial intelligence and in CEN/CENELEC JTC 21 on European AI standards --- is the space that the present contribution addresses.

\subsection{Methodological stance}

The paper adopts a normative-reconstructive methodology. It starts from the obligations that AI regulation imposes on the actors identified by such regulation --- in the EU context, the operators defined in Article 3(8) of the AI Act, whose categorization has been analyzed in \cite{fabiano2025subjects}; in other jurisdictions, the analogous categories of developers, deployers, and users defined by the relevant regulatory instruments --- and asks, from the perspective of each obligation, what kind of professional figure is in a position to interpret and apply it with the required precision. The approach builds on a methodological tradition of structural relationship modeling in data protection, in which the articulation of obligations is grounded in the relations among actors rather than in the isolated description of each \cite{fabiano2020dappremo, fabiano2024dappremo}.

The approach adopted here is \textit{juridical-first}, in deliberate distinction from three alternative orientations: technology-first approaches, which derive professional profiles from the architecture of the systems to be governed; market-first approaches, which derive them from the segmentation of existing consultancy offerings; and compliance-first approaches, which derive them from the operational workflows of existing compliance functions. Each of these alternatives has its own legitimacy, but none of them places the interpretation of the regulatory obligation at the center of the professional figure. The juridical-first approach, by contrast, asks what professional figure is required by the obligations themselves, and derives the profile from that requirement. This approach is intentionally jurisdiction-agnostic at the level of conceptual architecture, while being jurisdiction-specific at the level of regulatory content.

The reference to the European e-Competence Framework is adopted as a methodological choice, not as a formal endorsement of any specific certification scheme. The e-CF \cite{cenws2019} provides a European-recognized vocabulary of competences, with verifiable levels and an articulation that makes it possible to discuss professional profiles in commensurable terms. Alternative frameworks --- DigComp, ESCO, NIST AI Risk Management Framework \cite{nistairmf}, and the SFIA framework used in several Commonwealth jurisdictions --- are complementary but do not offer the same granularity for competence mapping at the level of the individual professional. The choice of the e-CF as reference does not imply that the profile is European in scope: the framework is used as a commensurable vocabulary, and the resulting competence architecture is transferable to alternative frameworks in other jurisdictions.

\subsection{Related work}
\label{sec:related}

The positioning of the AI Legal Specialist profile benefits from explicit comparison with the existing landscape of international standards, institutional guidance, and scientific contributions on AI governance and compliance.

On the standardization side, ISO/IEC 42001:2023 defines requirements for artificial intelligence management systems at the organizational level \cite{iso42001}, and ISO/IEC 38507:2022 addresses the governance implications of organizations' use of AI \cite{iso38507}. Both standards require organizations to define roles and responsibilities for AI governance, but leave the professional articulation of such roles outside their scope. The NIST AI Risk Management Framework \cite{nistairmf} provides a voluntary framework for AI risk governance widely referenced in the United States and beyond, but it does not define professional profiles. At the Council of Europe level, the Framework Convention on Artificial Intelligence \cite{coeaiconv2024} establishes the first binding international instrument on AI; its implementation will require professional figures capable of operationalizing the convention's obligations at the national level across signatory jurisdictions.

On the scientific side, the literature at the intersection of AI and law has predominantly addressed the application of AI to legal practice --- legal interpretation through large language models, automated legal reasoning, and specialized legal AI systems. The reverse direction --- the professional figure of the jurist specialized in AI as a regulated object --- has received substantially less attention. Recent work on the subject roles of the AI Act \cite{fabiano2025subjects}, on the critical oversight of large language models \cite{fabiano2024aillm}, and on the regulatory treatment of affective computing \cite{fabiano2025affective} contributes to this direction but does not address the definition of a dedicated professional profile. The broader discussion on the balance among innovation, knowledge, and ethics in the digital age \cite{fabiano2025aibook} and on the interaction among robotics, ethics, and data protection \cite{fabiano2019robotics} provides the substantive background against which the profile is defined.

The gap that the present paper addresses is therefore specific: a juridically grounded definition of the professional profile of the jurist specialized in AI, conceived as juridically autonomous, articulated through a reference competence architecture, operationalized through measurable indicators, and constructed to remain valid across the jurisdictions where AI regulation is being articulated. To the best of the author's knowledge, no prior scientific contribution has addressed the profile with this combination of elements.

\section{Defining the AI Legal Specialist}
\label{sec:autonomy}

\subsection{Mission and scope}

The AI Legal Specialist is a jurist whose professional function consists in the interpretation and application of the regulatory framework governing artificial intelligence systems, and in the consequent advisory and assurance activity directed to public and private organizations that develop, deploy, or use such systems. The figure operates at the structural intersection between legal knowledge and the governance of AI, and its distinguishing mark is the primacy of juridical reasoning as the interpretive key for the regulatory ecosystem of AI in the jurisdiction or jurisdictions of operation.

Three features define the profile at its core. First, juridical primacy: the specialist is, before any other attribute, a jurist --- not necessarily an enrolled advocate, but a professional with solid legal training and methodology, which may be exercised within the forensic profession, in academia, in public administration, in private organizations, or in independent research. The priority of the juridical dimension over the technical and managerial dimensions is not a claim of hierarchical superiority but a methodological observation: regulatory obligations, by their nature, can be authoritatively interpreted only through juridical methodology. A technologist with legal training addresses a different question from that of a jurist with technical literacy; both figures are legitimate and necessary, but the interpretation of the obligation itself remains, constitutively, a juridical act. Second, regulatory breadth: the subject-matter is not confined to a single AI regulation but encompasses the cluster of digital regulations that converge on AI systems within the relevant jurisdiction. In the European Union this cluster includes the AI Act, GDPR, NIS2, the Data Act, the DSA, the DMA, the Cyber Resilience Act, and the AI Liability regime; in the United States, the executive framework on AI, sector-specific instruments, state-level legislation on AI, and the data protection and consumer protection regimes; in other jurisdictions, the analogous combinations of AI-specific instruments, data protection law, cybersecurity law, and sectoral regulation. Third, technical literacy: the specialist possesses a working understanding of machine learning, deep learning, explainability techniques, bias and fairness, and data governance, sufficient for sustained dialogue with technical counterparts without external translation.

The three features are universal: they characterize the profile wherever it emerges. The specific content that populates each feature is, by contrast, jurisdiction-specific: the regulatory breadth varies with the regulatory bundle in force; the technical literacy is universally required but its specific applications depend on the technological systems prevalent in the jurisdiction's market; even the juridical primacy is exercised through the methodology of the legal tradition in which the jurist operates (civil law, common law, mixed systems). This combination of universal architecture and jurisdiction-specific content is the structural feature that distinguishes the AI Legal Specialist profile from both jurisdictionally limited profiles (such as those tied exclusively to the AI Act) and abstract technocratic profiles (such as those derived from technical standards alone).

\subsection{Autonomy with respect to adjacent figures}

The autonomy of the AI Legal Specialist can be articulated along four dimensions.

\textit{Autonomy with respect to the DPO and equivalent figures.} The mandate of the DPO, in the European context, and of analogous figures in jurisdictions whose data protection law derives from or converges with the GDPR, is defined by personal data processing; the AI Legal Specialist's mandate is defined by AI systems, whether or not they process personal data. The two profiles overlap only where AI systems process personal data, and in that intersection they work in coordination, not in substitution.

\textit{Autonomy with respect to the privacy lawyer.} The privacy lawyer is organized around data protection; the AI Legal Specialist is organized around AI governance. The first is defined by an object (personal data); the second is defined by a technology (AI systems). The objects intersect but do not coincide.

\textit{Autonomy with respect to the compliance officer.} The compliance officer produces organizational conformity; the AI Legal Specialist produces juridical interpretation as the basis for conformity. The first operates downstream of the legal question; the second operates upstream of it. In mature organizations the two figures are complementary and report to distinct lines of accountability.

\textit{Autonomy with respect to the generalist AI Governance professional.} The AI Governance professional, as sketched in recent certification schemes across jurisdictions, is a managerial figure whose legal competence is one dimension among several. The AI Legal Specialist is, conversely, a juridical figure whose competence in governance is instrumental to the juridical function. The difference is not semantic: it carries consequences for the reporting line, for the training pathway, and for the professional responsibility regime applicable to the figure.

It is occasionally suggested that the AI Legal Specialist profile might overlap with the DPO function, or be configured as a weak complement to it, in the sense of mere coexistence. Neither characterization is correct. The profile is not an extension of the DPO, nor a specialization of it, nor an auxiliary to it. It is a distinct figure with a distinct mandate, a distinct object of action, and a distinct methodological center of gravity. In organizations where both figures are present, they cooperate as autonomous professionals sharing the same institutional context, not as a primary and a secondary role within a single function.

\subsection{Knowledge and skill architecture}

The knowledge base of the AI Legal Specialist is structured around three clusters, each of which adapts to the jurisdiction of operation while preserving a common architecture. The first cluster is normative: the relevant AI-specific regulations together with the broader digital regulatory environment (data protection, cybersecurity, sectoral regulation, intellectual property), the soft-law and guidance instruments issued by competent authorities (in the European context, the European Data Protection Board, the AI Office, ENISA, and the Council of Europe; in other jurisdictions, the analogous bodies), and the case law and administrative practice developed under such instruments. The second cluster is technical: the functional understanding of AI systems, their development lifecycle, their failure modes, and their evaluation. The third cluster is methodological: risk assessment (DPIA, FRIA, Algorithmic Impact Assessment, and analogous instruments), audit, policy drafting, governance design, and training.

The skill base is correspondingly structured. Interpretive skills (statutory interpretation, case-law analysis, soft-law integration) interact with operational skills (drafting, negotiation, audit, training), and both rest on communicative skills that enable productive dialogue with technical, managerial, and institutional counterparts.

\section{Reference Competence Architecture}
\label{sec:ecf}

The competences required for the AI Legal Specialist profile can be described, at a first level of articulation, along four functional dimensions: strategic alignment (integration of AI regulatory requirements into organizational strategy); advisory on AI system design and verification (legal support to the development, integration, and testing of AI systems); support to AI service operation (legal assistance to AI systems in production and to their institutional interface); and process improvement for regulatory compliance (iterative refinement of AI compliance procedures in response to regulatory evolution). Each dimension corresponds to a cluster of competences that can be aligned with established competence frameworks. The present paper aligns the architecture with the European e-Competence Framework (e-CF), formalized in EN 16234-1 \cite{cenws2019}, as a methodological choice; equivalent alignments with alternative frameworks (such as SFIA in Commonwealth jurisdictions, or jurisdiction-specific competence taxonomies) are equally feasible without altering the conceptual architecture of the profile.

The alignment with the e-CF is a methodological choice of the profile's author, not a recommendation for third parties. The complete articulation of the alignment --- selection of specific competences, calibration of proficiency levels, and justification of each choice --- is set out in the deposited reference document, the full text of which remains reserved at the present stage. This paper discusses the conceptual architecture and the methodological justification of the alignment, but does not reproduce its specific elements. The reason for this reservation is substantive: the value of a competence architecture depends on the coherence of the entire system, and partial disclosure risks generating derivative profiles that would inherit the structure without the reasoning that justifies it. The author's intent is to preserve the integrity of the profile definition until it can be presented in an institutional context adequate to its nature. This reservation is consistent with the certified timestamp of the reference document (eIDAS, 11 May 2025) and with the author's commitment to contribute the full formalization to appropriate standardization or institutional venues in due course.

The general principle underlying the level calibration can nonetheless be stated. Competences of a strategic and policy-shaping character --- those that concern the alignment of the organization with the regulatory framework and the design of governance architectures --- attract the higher levels of proficiency envisaged by the framework, reflecting the strategic character of the function exercised by the profile. Competences of an operational and execution-oriented character --- those that concern the support to specific development and verification activities and to the institutional interface --- attract intermediate levels. The pattern is coherent with the juridical primacy of the profile: the jurist specialist shapes strategy and policy at the highest level of abstraction, and supports operational implementation at the level of institutional and procedural design, without substituting for the technical execution carried out by other roles.

This reference architecture is intended as a conceptual template, not as a prescriptive scheme. Organizational and jurisdictional contexts may justify variations in the selection of competences and in the calibration of levels; the value of the reference architecture lies in providing a commensurable starting point for discussion, training design, and certification alignment, while the full specification of the architecture remains reserved in the terms set out above.

\section{Operationalization through Key Performance Indicators}
\label{sec:kpi}

A professional profile that cannot be measured risks reducing itself to declaratory rhetoric. The operational entry of the AI Legal Specialist into organizational processes requires indicators that allow the verification, ex post, of the contribution delivered. The paper proposes six areas of measurement, which cover the complete lifecycle of AI legal compliance within an organization: regulatory input (area one), risk governance (area two), contractual formalization (area three), institutional embedding (area four), organizational diffusion (area five), and external interface (area six). Alternative segmentations are possible; the structure adopted here is justified by its coverage of both the internal and external dimensions of the function, and by its compatibility with existing performance management systems across jurisdictions. Each area is articulated around a cluster of indicators whose structural logic --- rather than their specific formulation --- is presented here.

The first area concerns regulatory conformity: the alignment of the organization's AI systems with the applicable regulatory framework, measured through indicators such as the outcome of legal audits, the update frequency of internal AI policies relative to regulatory evolution, the average time of adaptation to new regulatory requirements, and the number of non-conformity notices received from supervisory authorities.

The second area concerns risk management and legal mitigation: the identification, assessment, and mitigation of legal risks associated with AI systems, measured through indicators such as the number of AI impact assessments conducted, the percentage of systems aligned with explainability and fairness requirements, the number of bias or discrimination cases identified and corrected, and the number of risk management procedures implemented.

The third area concerns contractual and documentary management: the quality and currency of the contractual infrastructure governing AI procurement and deployment, measured through indicators such as the number of AI contracts drafted or revised, the average revision time, the update rate of contractual clauses relative to regulatory evolution, and the coverage of AI-specific terms in service agreements and privacy policies.

The fourth area concerns AI governance and ethics: the institutional infrastructure of the organization's AI function, measured through indicators such as the number of AI policies created or updated, the implementation rate of ethical guidelines, the frequency of ethical audits, and the average response time to ethical signals from internal and external stakeholders.

The fifth area concerns training and awareness: the diffusion of AI legal and compliance literacy within the organization, measured through indicators such as the number of training sessions delivered, the percentage of personnel trained, the qualitative feedback on training activities, and the number of guidance materials produced.

The sixth area concerns relations with stakeholders and regulatory authorities: the interface function that the profile exercises toward the external regulatory environment, measured through indicators such as the rate of successful regulatory audits, the number of compliance notices managed, the frequency of participation in institutional working groups, and the rate of closure of regulatory inquiries without enforcement action.

These six areas, taken together, provide a measurement surface sufficient for the integration of the profile into organizational performance management systems and into professional certification schemes. The indicators are intentionally formulated at the level of structure rather than of specific thresholds, which must be calibrated to the organizational and jurisdictional context.

\section{Discussion}

\subsection{Autonomy of the profile and its position in the existing landscape}

The juridical autonomy of the profile, affirmed at the opening of the present contribution, entails a consequence that deserves to be made explicit: the figure of the AI Legal Specialist does not presently find a place within the technical standardization frameworks currently available in the field of artificial intelligence. Instruments such as ISO/IEC 42001:2023 on AI management systems, ISO/IEC 38507:2022 on the governance of AI use by organizations \cite{iso38507}, and the NIST AI Risk Management Framework \cite{nistairmf} articulate organizational, procedural, and risk-management architectures of recognized value, but they do not undertake the definition of professional figures, since this is not their object. The observation is descriptive and does not imply any evaluation of such instruments, whose merits within their respective domains remain intact.

It follows that the figure of the jurist specialized in artificial intelligence represents, in the current global landscape, a professional space that is substantially unexplored from the perspective of its formalization. This condition is neither a shortcoming nor a problem in itself: it simply reflects the circumstance that the definition of a dedicated juridical figure belongs to a different order of discourse from that of technical standardization and organizational guidance. The jurist is defined through the structure of the obligations that he or she is called upon to interpret, not through the architecture of the processes in which the jurist operates.

The definition of such a profile, once articulated, yields benefits for the entire AI governance sector considered as a whole, across jurisdictions. These benefits manifest themselves on several coordinated planes. On the plane of professional practice, public and private organizations gain access to a recognizable figure to whom they may turn for the authoritative interpretation of obligations, reducing the applicative uncertainty that characterizes the current phase. On the plane of legal education, law schools and professional associations gain a conceptual reference around which to organize dedicated curricular pathways. On the plane of organizational adaptation, enterprises and public administrations adopting artificial intelligence systems can allocate juridical-interpretive responsibilities with greater precision, distinguishing them from operational management and procedural oversight responsibilities. On the plane of interinstitutional dialogue, the presence of a professional figure that can be identified by name facilitates communication among legal practitioners, technical experts, organizational decision-makers, and supervisory authorities. Each of these planes represents an autonomous contribution to the maturation of the sector, neither derived from nor dependent upon any specific instrument of standardization.

\subsection{The international dimension of AI regulation}

The profile is not jurisdictionally bounded. The Council of Europe Framework Convention on Artificial Intelligence \cite{coeaiconv2024}, adopted in May 2024 and opened for signature in September 2024, establishes binding obligations for signatory States including those outside the European Union. Its implementation at the national level will require professional figures able to translate its obligations into domestic compliance architectures. The profile articulated in this paper, with its three constitutive features --- juridical primacy, regulatory breadth, technical literacy --- is directly applicable to the implementation of the Convention in any signatory jurisdiction.

Beyond the Council of Europe instrument, the profile is relevant wherever AI is becoming the object of substantive regulation. In the United States, the combination of the executive framework, sector-specific federal regulation, and state-level AI legislation generates an obligation environment for which the profile's three features are well suited; the regulatory breadth in this context includes AI-specific federal and state instruments, the data protection and consumer protection regimes, and sectoral regulation in finance, health, and critical infrastructure. In the United Kingdom, the pro-innovation approach distributed across existing regulators generates a different but no less demanding obligation environment, in which the profile operates at the interface between AI-specific guidance and the regulatory mandates of the Information Commissioner's Office, the Competition and Markets Authority, the Financial Conduct Authority, and other bodies. In Canada, the Artificial Intelligence and Data Act proposal, complemented by federal and provincial privacy law and sector-specific regulation, defines the profile's operational domain. In Brazil, China, Japan, the Republic of Korea, Singapore, and other jurisdictions, the analogous combinations of AI-specific instruments and pre-existing regulatory frameworks define the corresponding operational domains.

The conceptual architecture of the profile is thus universal, while its specific operational content is jurisdiction-specific. This combination is not accidental: it reflects the fact that the profile is constituted by the structure of regulatory obligations, which is universal in its logical form (rules requiring authoritative interpretation), and by the substantive content of those obligations, which varies with the jurisdiction. A jurist who has fully internalized the conceptual architecture of the profile in one jurisdiction can transition to the practice of the same profile in another jurisdiction by acquiring the substantive regulatory content of the new context, without need to reconstitute the underlying methodological framework.

\subsection{Implications for legal education}

The profile bears on the evolution of legal education in any jurisdiction where it emerges. The training pathway that prepares an AI Legal Specialist is not exhausted by a traditional legal education supplemented with a course on the locally applicable AI regulation. It requires sustained exposure to the technical literature on AI, to the methodological vocabulary of risk assessment, and to the institutional practice of international regulatory communities. Law schools and professional associations that take seriously the emergence of this profile will need to redesign parts of their curricula accordingly, adapting the substantive content to the jurisdiction while preserving the architectural elements common to the profile across jurisdictions.

\subsection{Relation to existing legal professions}

A further consideration concerns the relationship of the profile with existing legal professions. The AI Legal Specialist is not a substitute for the generalist lawyer nor a competitor of existing specializations. It is an additional figure whose appearance expands the menu of legal professions available to public and private organizations facing AI-related obligations. As with the emergence of the DPO in the European post-GDPR period, and with analogous figures emerging in other regulatory contexts (Privacy Officers in North America, Data Protection Officers in Latin America under regulatory schemes derived from or converging with the GDPR), the consolidation of the new figure will occur gradually and through the interplay of regulatory pressure, market demand, and institutional recognition. The pace of consolidation will vary across jurisdictions, but the underlying logic of emergence is the same.

\section{Conclusions}

The acceleration of AI regulation across jurisdictions has generated a global demand for legal expertise that existing professional categories address only in adaptive form. This paper has argued that the adaptive response is structurally insufficient, and has proposed a juridically grounded and juridically autonomous definition of a distinct profile, the AI Legal Specialist, articulated around three features: juridical primacy, regulatory breadth, and technical literacy. The profile has been positioned with respect to adjacent figures and to international standardization instruments; its reference competence architecture has been aligned with the European e-Competence Framework as a methodological choice, while reserving the full articulation of its specific elements to the deposited reference document; and it has been operationalized through six areas of measurement whose implementation in organizational processes is a condition for the profile to move beyond declaratory existence.

The profile has been constructed to remain valid across jurisdictions: its conceptual architecture is universal, while the substantive regulatory content that populates its operational domain is jurisdiction-specific. The EU AI Act, as the most comprehensive and mature instance of horizontal AI regulation, provides the most articulated reference for the profile's regulatory breadth, but the profile's existence is not contingent on the AI Act alone: it emerges, with the same architectural features, in every jurisdiction where AI becomes the object of substantive regulation, including jurisdictions that implement the Council of Europe Framework Convention and those that develop autonomous regulatory approaches.

The paper is offered as a foundational contribution to a discussion that will accelerate in the coming years, as AI regulation matures across jurisdictions, as adjacent regulatory instruments enter into force, as the Council of Europe Framework Convention is implemented at the national level, and as international coordination on AI governance progresses. The definition proposed here is not final: it is a reference architecture, intended to be refined through dialogue with professional associations, standardization bodies, academic institutions, and the international scientific community across jurisdictions. The author invites engagement on the proposal and its operational development.


\begin{thebibliography}{99}

\bibitem{aiact2024}
European Parliament and Council. \textit{Regulation (EU) 2024/1689 of the European Parliament and of the Council of 13 June 2024 laying down harmonized rules on artificial intelligence (Artificial Intelligence Act)}. Official Journal of the European Union, L series, 12 July 2024. ELI: \url{http://data.europa.eu/eli/reg/2024/1689/oj}.

\bibitem{gdpr2016}
European Parliament and Council. \textit{Regulation (EU) 2016/679 of the European Parliament and of the Council of 27 April 2016 on the protection of natural persons with regard to the processing of personal data and on the free movement of such data (General Data Protection Regulation)}. Official Journal of the European Union, L 119, 4 May 2016. ELI: \url{http://data.europa.eu/eli/reg/2016/679/oj}.

\bibitem{nis22022}
European Parliament and Council. \textit{Directive (EU) 2022/2555 of the European Parliament and of the Council of 14 December 2022 on measures for a high common level of cybersecurity across the Union (NIS2 Directive)}. Official Journal of the European Union, L 333, 27 December 2022. ELI: \url{http://data.europa.eu/eli/dir/2022/2555/oj}.

\bibitem{dataact2023}
European Parliament and Council. \textit{Regulation (EU) 2023/2854 of the European Parliament and of the Council of 13 December 2023 on harmonized rules on fair access to and use of data (Data Act)}. Official Journal of the European Union, L series, 22 December 2023. ELI: \url{http://data.europa.eu/eli/reg/2023/2854/oj}.

\bibitem{dsa2022}
European Parliament and Council. \textit{Regulation (EU) 2022/2065 of the European Parliament and of the Council of 19 October 2022 on a Single Market For Digital Services (Digital Services Act)}. Official Journal of the European Union, L 277, 27 October 2022. ELI: \url{http://data.europa.eu/eli/reg/2022/2065/oj}.

\bibitem{dma2022}
European Parliament and Council. \textit{Regulation (EU) 2022/1925 of the European Parliament and of the Council of 14 September 2022 on contestable and fair markets in the digital sector (Digital Markets Act)}. Official Journal of the European Union, L 265, 12 October 2022. ELI: \url{http://data.europa.eu/eli/reg/2022/1925/oj}.

\bibitem{coeaiconv2024}
Council of Europe. \textit{Framework Convention on Artificial Intelligence and Human Rights, Democracy and the Rule of Law}. CETS No. 225, adopted 17 May 2024, opened for signature 5 September 2024, Vilnius. Available at: \url{https://www.coe.int/en/web/artificial-intelligence/the-framework-convention-on-artificial-intelligence}.

\bibitem{cenws2019}
European Committee for Standardization (CEN). \textit{EN 16234-1:2019 e-Competence Framework (e-CF) --- A common European framework for ICT Professionals in all sectors --- Part 1: Framework}. CEN, Brussels, 2019.

\bibitem{iso42001}
International Organization for Standardization. \textit{ISO/IEC 42001:2023 Information technology --- Artificial intelligence --- Management system}. ISO, Geneva, 2023.

\bibitem{iso38507}
International Organization for Standardization. \textit{ISO/IEC 38507:2022 Information technology --- Governance of IT --- Governance implications of the use of artificial intelligence by organizations}. ISO, Geneva, 2022.

\bibitem{nistairmf}
National Institute of Standards and Technology (NIST). \textit{Artificial Intelligence Risk Management Framework (AI RMF 1.0)}. NIST AI 100-1, January 2023. Available at: \url{https://doi.org/10.6028/NIST.AI.100-1}.

\bibitem{fabiano2024aillm}
N.~Fabiano. \textit{AI Act and Large Language Models (LLMs): When critical issues and privacy impact require human and ethical oversight}. arXiv preprint arXiv:2404.00600, 2024. DOI: \url{https://doi.org/10.48550/arXiv.2404.00600}.

\bibitem{fabiano2025subjects}
N.~Fabiano. \textit{Subject Roles in the EU AI Act: Mapping and Regulatory Implications}. arXiv preprint arXiv:2510.13591, 2025. DOI: \url{https://doi.org/10.48550/arXiv.2510.13591}.

\bibitem{fabiano2025affective}
N.~Fabiano. \textit{Affective Computing and Emotional Data: Challenges and Implications in Privacy Regulations, The AI Act, and Ethics in Large Language Models}. arXiv preprint arXiv:2509.20153, 2025. DOI: \url{https://doi.org/10.48550/arXiv.2509.20153}.

\bibitem{fabiano2025aibook}
N.~Fabiano. \textit{Artificial Intelligence, Neural Networks and Privacy: Striking a Balance between Innovation, Knowledge, and Ethics in the Digital Age}. Forewords by Danilo P. Mandic and Carlo Morabito; introduction by Guido Scorza. goWare, Florence, 2025. ISBN 978-88-3363-684-4.

\bibitem{fabiano2020dappremo}
N.~Fabiano. \textit{GDPR \& Privacy: Awareness and Opportunities. The approach with the Data Protection and Privacy Relationships Model (DAPPREMO)}. Foreword by Wojciech R. Wiewiórowski. goWare, Florence, 2020. ISBN 978-88-3363-407-4.

\bibitem{fabiano2024dappremo}
N.~Fabiano. A Singular Approach to Address Privacy Issues by the Data Protection and Privacy Relationships Model (DAPPREMO). In: K.~Rannenberg, P.~Drogkaris, C.~Lauradoux (eds.), \textit{Privacy Technologies and Policy}. Proceedings of the Annual Privacy Forum (APF 2023). Lecture Notes in Computer Science, Springer Nature Switzerland, 2024, pp. 166--181. ISBN 978-3-031-61089-9. DOI: \url{https://doi.org/10.1007/978-3-031-61089-9_8}.

\bibitem{fabiano2019robotics}
N.~Fabiano. Robotics, Big Data, Ethics and Data Protection: A Matter of Approach. In: \textit{Robotics and Well-Being}. Springer, 2019. DOI: \url{https://doi.org/10.1007/978-3-030-12524-0_8}.

\end{thebibliography}
\end{document}